\documentclass[twocolumn,showkeys,showpacs,preprintnumbers,prd,superscriptaddress,nofootinbib,longbibliography]{revtex4-1}

\usepackage{graphicx}
\usepackage{epsf}
\usepackage[bookmarks=false]{hyperref}
\usepackage{bm}
\usepackage{amsmath}
\usepackage{amsfonts}
\usepackage{amssymb}
\usepackage{epstopdf}
\usepackage{natbib}
\usepackage{color}
\usepackage{verbatim}
\usepackage{multirow}
\usepackage{xcolor}
\usepackage{mathrsfs}
\usepackage{alphalph}
\usepackage[T1]{fontenc}
\usepackage[utf8]{inputenc}

\date{\today}

\begin{document}
	


\newcommand{\red}[1]{\textcolor{red}{#1}}
\newcommand{\blue}[1]{\textcolor{blue}{#1}}
\newcommand{\gray}[1]{\textcolor{gray}{#1}} 
\newcommand{\yiannis}[1]{\textcolor{blue!70}{[#1]}} 

\newcommand{\wei}[2]{\textcolor{blue}{#1 }\todo[color=green]{WS: #2}}
\newcommand{\pankaj}[2]{\textcolor{blue}{#1 }\todo[color=yellow]{PC: #2}}
\newcommand{\boyang}[2]{\textcolor{blue}{#1 }\todo[color=orange]{BY: #2}}
\newcommand{\ioannis}[2]{\textcolor{blue}{#1 }\todo[color=cyan]{IP: #2}}

\newcommand{\plan}[1]{\begin{flushleft}
\gray{\tt *** #1 ***}
\end{flushleft}}

\def\bal#1\eal{\begin{align}#1\end{align}}

\def \bib{\bibitem}
\def\){\right)}
\def\({\left( }
\def\]{\right] }
\def\[{\left[ }

\def\nn{\nonumber}
\def\NO{\nonumber}
\def\nonu{\nonumber \\}
\def\ni{\noindent}

\def\half{\frac{1}{2}}


\newtheorem{definition}{Definition}[section]
\newtheorem{theorem}{Theorem}[section]
\newtheorem{lemma}{Lemma}[section]
\newtheorem{corollary}{Corollary}[section]
\newtheorem{proposition}{Proposition}[section]
\newtheorem{conjecture}{Conjecture}[section]


\def\a{\alpha}
\def\b{\beta}
\def\c{\chi}
\def\d{\delta}
\def\e{\epsilon}
\def\f{\phi}
\def\g{\gamma}
\def\h{\eta}
\def\j{\psi}
\def\k{\kappa}
\def\l{\lambda}
\def\m{\mu}
\def\n{\nu}
\def\om{\omega}
\def\p{\pi}
\def\th{\theta}
\def\r{\rho}
\def\s{\sigma}
\def\t{\tau}
\def\x{\xi}
\def\z{\zeta}
\def\D{\Delta}
\def\F{\Phi}
\def\G{\Gamma}
\def\J{\Psi}
\def\L{\Lambda}
\def\Om{\Omega}
\def\P{\Pi}
\def\Th{\Theta}
\def\S{\Sigma}
\def\U{\Upsilon}
\def\X{\Xi}


\def\ve{\varepsilon}
\def\vr{\varrho}
\def\vs{\varsigma}
\def\vth{\vartheta}
\def\tvf{\tilde{\varphi}}
\def\vf{\varphi}


\def\bba{\bbalpha}
\def\bbk{\bbkappa}
\def\bbg{\bbgamma}
\def\bbd{\bbdelta}
\def\bbs{\bbsigma}


\def\ca{{\cal A}}
\def\cb{{\cal B}}
\def\cc{{\cal C}}
\def\cd{{\cal D}}
\def\ce{{\cal E}}
\def\cf{{\cal F}}
\def\cg{{\cal G}}
\def\ch{{\cal H}}
\def\ci{{\cal I}}
\def\cj{{\cal J}}
\def\ck{{\cal K}}
\def\cl{{\cal L}}
\def\cm{{\cal M}}
\def\cn{{\cal N}}
\def\co{{\cal O}}
\def\cp{{\cal P}}
\def\cq{{\cal Q}}
\def\car{{\cal R}}
\def\cs{{\cal S}}
\def\ct{{\cal T}}
\def\cu{{\cal U}}
\def\cv{{\cal V}}
\def\cw{{\cal W}}
\def\cx{{\cal X}}
\def\cy{{\cal Y}}
\def\cz{{\cal Z}}


\def\bta{\textit{\textbf{a}}}
\def\btb{\textit{\textbf{b}}}
\def\btc{\textit{\textbf{c}}}
\def\btd{\textit{\textbf{d}}}
\def\bte{\textit{\textbf{e}}}
\def\btf{\textit{\textbf{f}}}
\def\btg{\textit{\textbf{g}}}
\def\bth{\textit{\textbf{h}}}
\def\bti{\textit{\textbf{i}}}
\def\btj{\textit{\textbf{j}}}
\def\btk{\textit{\textbf{k}}}
\def\btl{\textit{\textbf{l}}}
\def\btm{\textit{\textbf{m}}}
\def\btn{\textit{\textbf{n}}}
\def\bto{\textit{\textbf{o}}}
\def\btp{\textit{\textbf{p}}}
\def\btq{\textit{\textbf{q}}}
\def\btr{\textit{\textbf{r}}}
\def\bts{\textit{\textbf{s}}}
\def\btt{\textit{\textbf{t}}}
\def\btu{\textit{\textbf{u}}}
\def\btv{\textit{\textbf{v}}}
\def\btw{\textit{\textbf{w}}}
\def\btx{\textit{\textbf{x}}}
\def\bty{\textit{\textbf{y}}}
\def\btz{\textit{\textbf{z}}}

\newcommand{\sgn}{\operatorname{sgn}}

\renewcommand{\bar}{\overline}
\renewcommand{\tilde}{\widetilde}
\renewcommand{\hat}{\widehat}
\renewcommand{\leq}{\leqslant}
\renewcommand{\geq}{\geqslant}
\newcommand{\la}{\left\langle}
\newcommand{\ra}{\right\rangle}
\newcommand{\xp}{x^{+}}
\newcommand{\xm}{x^{-}}

\newcommand{\CC}{\mathbb{C}}
\newcommand{\RR}{\mathbb{R}}
\newcommand{\HH}{\mathbb{H}}
\newcommand{\ZZ}{\mathbb{Z}}
\newcommand{\cA}{\mathcal{A}}
\newcommand{\cB}{\mathcal{B}}
\newcommand{\cC}{\mathcal{C}}
\newcommand{\cD}{\mathcal{D}}
\newcommand{\cE}{\mathcal{E}}
\newcommand{\cF}{\mathcal{F}}
\newcommand{\cG}{\mathcal{G}}
\newcommand{\cH}{\mathcal{H}}
\newcommand{\cI}{\mathcal{I}}
\newcommand{\cJ}{\mathcal{J}}
\newcommand{\cK}{\mathcal{K}}
\newcommand{\cL}{\mathcal{L}}
\newcommand{\cM}{\mathcal{M}}
\newcommand{\cN}{\mathcal{N}}
\newcommand{\cO}{\mathcal{O}}
\newcommand{\cP}{\mathcal{P}}
\newcommand{\cQ}{\mathcal{Q}}
\newcommand{\cR}{\mathcal{R}}
\newcommand{\cS}{\mathcal{S}}
\newcommand{\cT}{\mathcal{T}}
\newcommand{\cU}{\mathcal{U}}
\newcommand{\cV}{\mathcal{V}}
\newcommand{\cW}{\mathcal{W}}
\newcommand{\cX}{\mathcal{X}}
\newcommand{\cY}{\mathcal{Y}}
\newcommand{\cZ}{\mathcal{Z}}


\newcommand{\bra}{\bar{a}}
\newcommand{\brb}{\bar{b}}
\newcommand{\brc}{\bar{c}}
\newcommand{\brd}{\bar{d}}
\newcommand{\bre}{\bar{e}}
\newcommand{\brf}{\bar{f}}
\newcommand{\brg}{\bar{g}}
\newcommand{\brh}{\bar{h}}
\newcommand{\bri}{\bar{i}}
\newcommand{\brj}{\bar{j}}
\newcommand{\brk}{\bar{k}}
\newcommand{\brl}{\bar{l}}
\newcommand{\brm}{\bar{m}}
\newcommand{\brn}{\bar{n}}
\newcommand{\bro}{\bar{o}}
\newcommand{\brp}{\bar{p}}
\newcommand{\brq}{\bar{q}}
\newcommand{\brr}{\bar{r}}
\newcommand{\brs}{\bar{s}}
\newcommand{\brt}{\bar{t}}
\newcommand{\bru}{\bar{u}}
\newcommand{\brv}{\bar{v}}
\newcommand{\brw}{\bar{w}}
\newcommand{\brx}{\bar{x}}
\newcommand{\bry}{\bar{y}}
\newcommand{\brz}{\bar{z}}

\newcommand{\be}{\begin{equation}}
\newcommand{\ee}{\end{equation}}
\newcommand{\bea}{\begin{eqnarray}}
\newcommand{\eea}{\end{eqnarray}}
\newcommand{\bb}{\mathbb}
\newcommand{\ba}{\begin{align}}
\newcommand{\ea}{\end{align}}
\newcommand{\bad}{\begin{aligned}}
\newcommand{\ead}{\end{aligned}}
\newcommand{\nd}{\noindent}
\newcommand{\bsub}{\begin{subequations}}
\newcommand{\esub}{\end{subequations}}
\newcommand{\beqx}{\begin{displaymath}}
\newcommand{\eeqx}{\end{displaymath}}
\newcommand{\bmat}{\left(\begin{array}}
\newcommand{\emat}{\end{array}\right)}
\newcommand*\Laplace{\mathop{}\!\mathbin\bigtriangleup}
\newcommand*\DAlambert{\mathop{}\!\mathbin\Box}



\def\Sc#1{{\hbox{\sc #1}}}      
\def\Sf#1{{\hbox{\sf #1}}}      
\def\mb#1{\mbox{\boldmath $#1$}}
\def\mf#1{\ensuremath{\mathfrak{#1}}} 
\def\bb#1{\ensuremath{\mathbb{#1}}} 


\def\slpa{\slash{\pa}}                         
\def\slin{\SLLash{\in}}                                 
\def\bo{{\raise-.3ex\hbox{\large$\Box$}}}               
\def\cbo{\Sc [}                                         
\def\pa{\partial}                                       
\def\de{\nabla}                                         
\def\dell{\nabla}                                       
\def\su{\sum}                                           
\def\pr{\prod}                                          
\def\iff{\leftrightarrow}                               
\def\conj{{\hbox{\large *}}}                            
\def\ltap{\raisebox{-.4ex}{\rlap{$\sim$}} \raisebox{.4ex}{$<$}}   
\def\gtap{\raisebox{-.4ex}{\rlap{$\sim$}} \raisebox{.4ex}{$>$}}   
\def\face{{\raise.2ex\hbox{$\displaystyle \bigodot$}\mskip-2.2mu \llap {$\ddot
        \smile$}}}                                   
\def\dg{\dagger}                                     
\def\ddg{\ddagger}                                   
\def\trans{\mbox{\scri T}}                           
\def\>{\rangle}                                      
\def\<{\langle}                                      


\def\tx#1{\text{#1}}
\def\sp#1{{}^{#1}}                                   
\def\sb#1{{}_{#1}}                                   
\def\sptx#1{{}^{\rm #1}}                           
\def\sbtx#1{{}_{\rm #1}}                           
\newcommand{\sub}[1]{\phantom{}_{(#1)}\phantom{}}    
\def\oldsl#1{\rlap/#1}                               
\def\slash#1{\rlap{\hbox{$\mskip 1 mu /$}}#1}        
\def\Slash#1{\rlap{\hbox{$\mskip 3 mu /$}}#1}        
\def\SLash#1{\rlap{\hbox{$\mskip 4.5 mu /$}}#1}      
\def\SLLash#1{\rlap{\hbox{$\mskip 6 mu /$}}#1}       
\def\wt#1{\widetilde{#1}}                            
\def\Hat#1{\widehat{#1}}                             
\def\lbar#1{\ensuremath{\overline{#1}}}              
\def\VEV#1{\left\langle #1\right\rangle}             
\def\abs#1{\left| #1\right|}                         
\def\leftrightarrowfill{$\mathsurround=0pt \mathord\leftarrow \mkern-6mu
        \cleaders\hbox{$\mkern-2mu \mathord- \mkern-2mu$}\hfill
        \mkern-6mu \mathord\rightarrow$}        
\def\dvec#1{\vbox{\ialign{##\crcr
        \leftrightarrowfill\crcr\noalign{\kern-1pt\nointerlineskip}
        $\hfil\displaystyle{#1}\hfil$\crcr}}}           
\def\dt#1{{\buildrel {\hbox{\LARGE .}} \over {#1}}}     
\def\dtt#1{{\buildrel \bullet \over {#1}}}              
\def\der#1{{\pa \over \pa {#1}}}                        
\def\fder#1{{\d \over \d {#1}}}                         
\def\tr{{\rm tr \,}}                                    
\def\Tr{{\rm Tr \,}}                                    
\def\diag{{\rm diag \,}}                                
\def\Re{{\rm Re\,}}                                     
\def\Im{{\rm Im\,}}                                     
\def\mrp{\mathrm{p}}

\def\partder#1#2{{\partial #1\over\partial #2}}        
\def\parvar#1#2{{\d #1\over \d #2}}                    
\def\secder#1#2#3{{\partial^2 #1\over\partial #2 \partial #3}}  
\def\on#1#2{\mathop{\null#2}\limits^{#1}}              
\def\bvec#1{\on\leftarrow{#1}}                         
\def\oover#1{\on\circ{#1}}                             


\def\Deq#1{\mbox{$D$=#1}}                               
\def\Neq#1{\mbox{$cn$=#1}}                              
\newcommand{\ampl}[2]{{\cal M}\left( #1 \to #2 \right)} 


\def\NPB#1#2#3{Nucl. Phys. B {\bf #1} (19#2) #3}
\def\PLB#1#2#3{Phys. Lett. B {\bf #1} (19#2) #3}
\def\PLBold#1#2#3{Phys. Lett. {\bf #1}B (19#2) #3}
\def\PRD#1#2#3{Phys. Rev. D {\bf #1} (19#2) #3}
\def\PRL#1#2#3{Phys. Rev. Lett. {\bf #1} (19#2) #3}
\def\PRT#1#2#3{Phys. Rep. {\bf #1} C (19#2) #3}
\def\MODA#1#2#3{Mod. Phys. Lett.  {\bf #1} (19#2) #3}


\def\norder{\raisebox{-.13cm}{\ensuremath{\circ}}\hspace{-.174cm}\raisebox{.13cm}{\ensuremath{\circ}}}
\def\bz{\bar{z}}
\def\bw{\bar{w}}
\def\-{\hphantom{-}}
\newcommand{\dd}{\mbox{d}}
\newcommand{\scr}{\scriptscriptstyle}
\newcommand{\scri}{\scriptsize}
\def\rand#1{\marginpar{\tiny #1}}               
\newcommand{\rstar}{\rand{\bf\large *}}
\newcommand{\rup}{\rand{$\uparrow$}}
\newcommand{\rdown}{\rand{$\downarrow$}}


	\preprint{APS-XXX}
	
	\title{Rotating black hole solutions for $f(R)$ gravity and Newman Janis Algorithm} 
	
	\author{Pankaj Chaturvedi}
	\email{cpankaj1@gmail.com}
	\affiliation{Department of Physics, Ariel University, Ariel 40700, Israel}%
	
	\author{Utkarsh Kumar}%
	\email{kumarutkarsh641@gmail.com}
	\affiliation{Department of Physics, Ariel University, Ariel 40700, Israel}
	
	\author{Udaykrishna Thattarampilly}%
	\email{uday7adat@gmail.com}
	\affiliation{Department of Physics, Ariel University, Ariel 40700, Israel}

    \author{Vishnu Kakkat}%
	\email{kakkav@unisa.ac.za}
	\affiliation{Department of Mathematical Sciences, Unisa}

	\date{\today}
	
	\begin{abstract}
		We show that the $f(R)$-gravity theories with constant Ricci scalar in the Jordan/Einstein frame can be described by Einstein or Einstein-Maxwell gravity with a cosmological term and a modified gravitational constant. We also propose a modified Newmann-Janis algorithm to obtain the rotating axisymmetric solutions for the Einstein/Einstein-Maxwell gravity with a cosmological constant. Using the duality between the two gravity theories we show that the stationary or static solutions for the Einstein/Einstein-Maxwell gravity with a cosmological constant will also be the solutions for the dual $f(R)$-gravity with constant Ricci scalar.
	\end{abstract}

\maketitle
	
	
 \section{Introduction}

The general theory of relativity (GR) is widely accepted as the fundamental theory of spacetime and gravity. Despite predicting numerous observational tests at large distances and late time scales (Infrared regime) to match the measurements from solar system tests, GR has gone through several challenges from the observational and theoretical viewpoints. Cosmological observations pertaining to cosmic microwave background (CMB), Type Ia supernovae and several others indicate that the Universe has undergone two phases of cosmic acceleration namely inflation and dark energy that occurred at early and late times respectively \cite{SupernovaSearchTeam:1998fmf,1979JETPL..30..682S,PhysRevD,Huterer:1998qv,Sahni:1999gb}. GR, in its original form, is unable to explain these phases of cosmic acceleration. Cosmological constant used to parametrize the recent accelerated expansion of the Universe is plagued by hierarchy problems in particle physics \cite{Weinberg:2000yb}.
Beyond its inconsistencies with cosmological and astrophysical data, GR also presents numerous theoretical weaknesses. Most prominently, GR struggles in the Ultraviolet spectrum, especially when interpreting the physics of black holes and cosmological singularities at short distances and small time intervals. This is due to the fact that GR is a 2-derivative action which poses problems of renormalizability at the quantum level. The non-renormalizability of GR has spurred interest in higher derivative gravity theories, often referred to as modified gravities.

 A simple yet consequential model for modified gravity is the
$f(R)$ gravity theory in which the Lagrangian density is modified to be an arbitrary analytic function of the Ricci scalar \cite{Bergmann1968CommentsOT,Ruzmaikina1969QuadraticCT}. The significance of higher-order terms in the $f(R)$ gravity Lagrangian can be attributed to the fact that they can descend from the low energy limit of String/M-theory \cite{Nojiri:2003rz}. The simplest $f(R)$-gravity with $ R^2$ correction can give rise to a phase of inflation which was first proposed and explored by Starobinsky\cite{Starobinsky:1980te}. There are numerous $f(R)$ gravity models to explain the cosmic inflation \cite{ElBourakadi:2023ufz,Wang:2023hsb,Ossoulian:2023moq,Belhaj:2023zmz,Oikonomou:2023bmn,Jeong:2023zrv,Odintsov:2023lbb,Oikonomou:2023qfz,Dioguardi:2023jwa,Cheraghchi:2023mjz,Odintsov:2023weg,Baffou:2023uki,Luongo:2023aaq} and current accelerated expansion of the Universe \cite{Capozziello:2002rd,Capozziello:2006ph,Capozziello:2003gx,Amendola:2006we,PhysRevD.75.084010,Amendola:2007nt,Appleby:2007vb,Starobinsky:2007hu,Cognola:2007zu}. In addition to that $f(R)$ gravity theories  also has been investigated in explaining the singularity problem arising in the strong gravity regime \cite{Astashenok:2022uhw,Nava-Callejas:2022pip,Panotopoulos:2018enj,Yu:2017uyd,Doneva:2017jop,Cikintoglu:2017jfh,Sussman:2017qya,Gao:2016rdu,Canate:2015dda,Berti:2015itd}, galaxy rotation curves \cite{Mohan:2022kvb,Parbin:2022nqh,Shabani:2022buw,Sharma:2019yix,Naik:2019moz,Naik:2018mtx,Matsakos:2016csq,Dey:2014gka,Capozziello:2013yha}, detection of gravitational waves \cite{Dimastrogiovanni:2022eir,Odintsov:2022hxu,Oikonomou:2022pdf,Alves:2022yea,Oikonomou:2022ijs,Narang:2022jkv,Inagaki:2023tjh,Liang:2023fmc,Khlopov:2023mpo,Dyadina:2023hrm,Ezquiaga:2023xfe,Alves:2023rxs} and many more.

The current era of gravitational wave astronomy has presented us with the possibility of investigating physics of extremely compact objects, such as black holes and neutron stars \cite{Yunes:2016jcc}. This has opened new prospects for observing and testing theoretical models in the strong gravity regime. As a platform for theories that explain cosmic acceleration and inflation, it is of paramount importance to explore and test $f(R)$ theories of modified gravity. The study of the properties of black hole solutions in these scenarios can provide strong gravity tests for these theories and may hint toward significant deviations from the GR. This knowledge of black hole space-time can be obtained from solutions to the field equations, which, although easy to find analytically in General Relativity (GR), is a non-trivial task in modified gravity theories. Since most astrophysical objects are considered to be spinning, there is an interest in finding rotating black hole solutions for $f(R)$ theories. Several black hole solutions in modified gravities with  or without matter have been studied previously. Outside of $f(R)$ theories \cite{Cvetic:2001bk,Cai:2001dz,Moon:2011hq,Capozziello:2009jg,Nzioki:2009av,Olmo:2006eh} such studies involve  finding the black holes solutions for Gauss-Bonnet Gravity, Lovelock, non-local, and other modified theories studied in \cite{Cai:2003kt,Matyjasek:2006fq,Kumar:2018chy,Kumar:2018pkb,Kumar:2018jjb,Kumar:2019uwi,Kumar:2019lzp}.

It is well known that  $f(R)$ gravity in Jordan frame can be recast as a non-minimally coupled scalar-tensor gravity theory in the Einstein frame by means of a conformal transformation \cite{Nojiri:2010wj,Capozziello:2011et}. Thus for a problem formulated for $f(R)$ gravity, the usual approach is to first solve the simpler field equations of motion in the Einstein frame and then use a conformal transformation to revert back to the Jordan frame.  Spherically symmetric black hole solutions of  $f(R)$-gravity theories with or without matter have been studied in the Einstein/Jordan frame in \cite{Mignemi:1991wa,Multamaki:2006zb}. Although this conformal transformation of higher order gravity to scalar-tensor gravity is in general plausible, it does not shed enough information about the physical relevance of these two theories in different frames. This discrepancy is related to the fact that using a conformal transformation to go from one frame to the other the stability of the solutions and their physical meaning can completely change. This leads to several physical constraints which has to be imposed on the form of the function $f(R)$ in order to have stable solutions in both the Einstein and the Jordan frame \cite{Bhattacharya:2015nda}. Despite the ambiguities mentioned above, it has been well establish that any $f(R)$ gravity with constant Ricci scalar is dual to the Einstein gravity with a cosmological and an effective  gravitational constants \cite{Tsujikawa:2008in,DeFelice:2010aj}. 

Motivated by the the above duality of $f(R)$-gravity in the Jordan frame to scalar-tensor gravity in the Einstein frame, in this work we have explored the description of constant curvature $f(R)$ gravity in the Einstein frame. We show that the constant curvature $f(R)$-gravity in the Einstein frame is indeed described by Einstein Gravity with a cosmological and an effective  gravitational constants at the level of the action and field equations. In this work, we also examine the possible static or stationary blackhole solutions for the constant curvature $f(R)$-gravity with Maxwell field in the Einstein frame.
 
In general it is not an easy task to obtain the rotating black hole solutions in the modified or the simple Einstein gravity models with various horizon topologies because of the nonlinear nature of the field equations. It was observed by Newman and Janis in the case of the Einstein equations that a certain algorithm produces the Kerr solution from the corresponding non-rotating counterpart. This procedure is now known as the Newman-Janis algorithm (NJA) \cite{newman1965metric,Adamo:2014baa}. This algorithm has been widely used as a technique to generate the Kerr-like rotating metrics from the corresponding static metrics, see \cite{Shao:2020weq,Kubiznak:2022vft,Kamenshchik:2023woo,Fernandes:2023vux,Ghosh:2023nkr}. This algorithm provides a way to generate axisymmetric metrics from a spherically symmetric stationary seed metric through a particular type of complexification of radial and time coordinates. Although this algorithm functions effectively within the framework of classical General Relativity (GR), the reason it yields the same result for the Kerr metric in classical GR remains somewhat elusive. Moreover, it has been illustrated in \cite{hansen2013applicability,Ferraro:2013oua,Ayzenberg:2014aka} that the NJ algorithm is not suitable for generating axisymmetric metrics in quadratic gravity models.  Furthermore, It is still unknown about the applicability of the NJ algorithm to the other modified gravity models. In this paper, we propose a modified version of the NJ algorithm in order to generate stationary axisymmetric blackhole solutions for various constant curvature $f(R)$ gravity models in the Einstein frame.
	
 The article is organised as following. In section \ref{sec:Jordan}, we review the Jordan and Einstein Frames for modified gravity and establish the duality between the constant curvature $f(R)$ gravity theories in Jordan frame to Einstein-Maxwell gravity with a CC. Furthermore, we explicitly show this duality for well-known examples. Section \ref{sec:modiNJ} deals with the modification in Newmann-Janis Algorithm to generate the rotating black hole solution for the Einstein gravity with CC. Finally, we discuss our findings and draw conclusions in section \ref{sec:Discussion}.

\section{$f(R)$ gravity in Jordan and Einstein frames} \label{sec:Jordan}
	The standard form of the action for $f(R)$ gravity in the so-called Jordan frame is given by 
	\begin{equation}
		{\cal S}_{\cJ}=\frac{1}{2\kappa}\int d^4x\sqrt{-g}\;f(R)+{\cal S}_M,\label{SJframe}
	\end{equation}
	where $\kappa=8\pi G$ in natural units with $G$ being the four dimensional gravitational constant, $f(R)$ is a generic function of the Ricci scalar $R$, and ${\cal S}_M$ is the usual matter contribution to the action. Varying the above action with respect to the metric results in the following Euler-Lagrange 
	equations of motion
	\begin{equation}
		\left(g_{ab}\Box-\nabla_a\nabla_b\right)F(R)+F(R)R_{ab}-\frac{1}{2}f(R) g_{ab}=\kappa T_{ab},\label{SJeom}
	\end{equation}
	where $F(R)=f'(R)$ and the matter stress-energy tensor $T_{ab}$ is given by
	\begin{equation}
		T_{ab}=-\frac{2}{\sqrt{-g}}\frac{\delta {\cal S}_{M}}{\delta g_{ab}}
	\end{equation}
	
	The following action describes the Einstein-Hilbert gravity with a non-minimally coupled scalar field
	\bal 
	\cS_{\cJ}=\frac{1}{2\kappa}\int d^4x\sqrt{-g}\left(F(\phi)R-V_{\cJ}(\phi)\right)+{\cal S}_M,\label{SJphiframe}
	\eal 
	where
	\bal 
	V_{\cJ}(\phi)=\phi F(\phi)-f(\phi),
	\eal
	is dynamically equivalent to the $f(R)$ gravity action in \eqref{SJframe}. This may be seen by considering the variation of the action in \eqref{SJphiframe} with respect to the metric and to the scalar field-$\phi(x_{\mu})$ which gives us the following Euler-Lagrange equations of motion
	\bal
	&\left(g_{ab}\Box-\nabla_a\nabla_b\right)F(\phi)+F(\phi)R_{ab}-\frac{1}{2}f(\phi) g_{ab}=\kappa T_{ab},\nn\\
	&F'(\phi) (R-\phi)-F(\phi)+f'(\phi)=0.\label{SJphieom}
	\eal
	
	Now provided $F'(\phi)\neq 0$ and $F(\phi)=f'(\phi)$ one gets the constraint $R=\phi$ from the second equation of motion in \eqref{SJphieom}. Plugging back the constraint $R=\phi$ in the action \eqref{SJphiframe} and the equations of motion \eqref{SJphieom} one recovers the $f(R)$ gravity action and equations of motion. The action in \eqref{SJphiframe} can now be recast in the Einstein frame by considering the following conformal transformation
	\bal
	\tilde{g}_{ab}=F(\phi)\;g_{ab},\label{cnftrans}
	\eal
	where one must impose $F(\phi)>0$ for the regularity of the given transformation. Given the new metric $\tilde{g}_{ab}$, the action for the $f(R)$ gravity can now be written in the Einstein frame as
	\bal
	&\cS_{E}=\int d^4x\sqrt{-\tilde{g}}\left(\frac{\tilde{R}}{2\kappa} -\frac{1}{2}\tilde{\nabla}_{a}\tilde{\phi}\;\tilde{\nabla}_{b}\tilde{\phi}-V_{E}(\tilde{\phi})\right)+\tilde{\cS}_{M},\label{SEframe}\\
	&\tilde{\phi}=\sqrt{\frac{3}{2\kappa}}\ln{\[F(\phi)\]},\quad V_{E}(\tilde{\phi})=\frac{V_{\cJ}(\phi)}{F(\phi)^2},\label{VphiVEexpr}\\
	&\tilde{R}=\frac{1}{F(\phi)}\left(R-3\frac{\Box F(\phi)}{F(\phi)}+\frac{3}{2}\frac{\nabla_a F(\phi)\nabla^a F(\phi)}{F(\phi)^2}\right),\label{REframe}
	\eal
	where the quantities with the subscript $\tilde{()}$ are defined with respect to the new $\tilde{g}_{ab}$-metric. Moreover, under the transformation \eqref{cnftrans} the matter stress-energy tensor transforms as
	\bal
	\tilde{T}_{ab}=\frac{T_{ab}}{F(\phi)^2}
	\eal
	
	The conformal equivalence of $f(R)$ gravity in the two different frames as described above, must be accompanied by certain consistency conditions. In particular, for the case when the form of $f(R)$ is described by a polynomial function of order greater than two, the correspondence between the two conformal frames becomes many-to-one \cite{Bhattacharya:2015nda}. This implies that there are multiple Einstein frame descriptions of a single higher-order $f(R)$ theory. As described in the introduction, for a given higher-order $f(R)$-gravity in the Jordan frame one has to consider $F'(R)>0$ and $F(R)>0$ as the necessary conditions for the existence of the corresponding Einstein frame. These conditions are also required to ensure the existence of a matter-dominated era in cosmological evolution in a high curvature classical regime, as elucidated in \cite{Bhattacharya:2015nda}. This motivates us to first consider the possible solutions of the $f(R)$ gravity in the Einstein frame. We also interpret the equivalence of such solutions in the Einstein frame with those in the Jordan frame.
	
	\subsection{ Constant curvature black hole solutions for $f(R)$ gravity in Einstein frame}
	
	The action in \eqref{SEframe} describes the $f(R)$ gravity in the Einstein frame. Considering the matter contribution coming solely from the Maxwell field ($A_{a}$) i.e.,
	\bal
	\tilde{S}_{M}=-\frac{1}{8\kappa}\int d^4x\sqrt{-\tilde{g}}\;\tilde{\cF}^2,\label{matterSEframe}
	\eal
	where $\tilde{\cF}^2=\tilde{\cF}_{ab}\tilde{\cF}^{ab}$ and $\tilde{\cF}_{ab}=\tilde{\nabla}_{a} A_{b}-\tilde{\nabla}_{b}A_{a}$ is the electromagnetic field strength tensor \footnote{The electromagnetic field strength tensor in the Einstein frame is related to the one in the Jordan frame by the transformation, {$\tilde{\cF}_{ab}=\cF_{ab}/F(\phi)^2$}.}, one can see that the action in \eqref{SEframe} describes an Einstein-Maxwell (EM) gravity non-minimally coupled to a scalar field. The corresponding Euler-Lagrange equations of the motion for the field $\tilde{\phi}$, $\tilde{g}_{ab}$, and $A_{a}$ can now be given as
	\bal
	&\tilde{R}_{ab}-\frac{1}{2}\tilde{g}_{ab}\tilde{R}-\left(\tilde{\nabla}_{a}\tilde{\phi}\;\tilde{\nabla}_{b}\tilde{\phi}-\tilde{g}_{ab}V_{E}(\tilde{\phi})\right.\nn\\
	&\left.+\tilde{\cF}_{ac}\tilde{\cF}^{cb}-\frac{1}{4}\tilde{g}_{ab}\tilde{\cF}^2\)=0,\label{Eframegeom}\\
	&\tilde{\nabla}_a\tilde{\nabla}^{a}\tilde{\phi}-\frac{\d V_{E}(\tilde{\phi})}{\d \tilde{\phi}}=0,\label{Eframephieom}\\
	&\tilde{\nabla}_{a}\(\tilde{\cF}^{ab}\)=0.\label{EframeAeom}
	\eal
	
	Determining a general solution for the above equations of motion is in general difficult when the scalar field $\tilde{\phi}$ has a dynamical (coordinate dependent) solution \cite{Cadoni:2011kv, Cadoni:2009xm, Martinez:2004nb, Gao:2004tu, Cadoni:2011nq, Serra_2012}. However, in the case when the scalar field $\tilde{\phi}$ has a constant profile then assuming, 
	\bal
	\tilde{\phi}=\cC,\quad F(\phi)=e^{\sqrt{\frac{2\kappa}{3}}\cC},\quad  V_{E}(\tilde{\phi})=\frac{\Lambda }{\kappa} e^{-\sqrt{\frac{2\kappa}{3}}\cC},\label{constphiVe}
	\eal
	where $\cC$ and $\Lambda$ are some constants, it may be observed that the action in \eqref{SEframe} reduces to 
	\bal
	\cS_{E}=&\frac{1}{{2\kappa}}\int d^4x\sqrt{-\tilde{g}}\left(\tilde{R}-2\Lambda e^{-\sqrt{\frac{2\kappa}{3}}\cC} -\frac{1}{4}\tilde{\cF}^2\right),\nn\\
	=& \frac{1}{2\tilde{\kappa}}\int d^4x\sqrt{-g}\left(R-2\Lambda -\frac{1}{4}\cF^2\right),\label{SEframeL}
	\eal
	which describes the Einstein-Maxwell (EM) gravity with a cosmological constant ($\Lambda$) and a modified effective gravitational constant ($G_{eff}$) given by
	\bal
	G_{eff}=\frac{\tilde{\kappa}}{8\pi}=G e^{-\sqrt{\frac{2\kappa}{3}}\cC}.\label{Geff} 
	\eal
	
	The cosmological constant is related to the AdS or dS length ($L$) as  $\Lambda=-\frac{3}{L^2}$ or  $\Lambda=\frac{3}{L^2}$ respectively. Depending on the sign of the cosmological constant, it is well known that the EM-gravity with a cosmological constant possesses several solutions namely, Reissner-Nordstr$\ddot{\text{o}}$m AdS/dS black hole and Kerr-Newmann AdS/dS black hole. Moreover, for vanishing Maxwell field these solutions reduce to Scwarzschild AdS/dS black hole and Kerr AdS/dS black hole respectively. These black holes usually belong to the constant $\tilde{R}$ ($\tilde{R}=4\Lambda$) solution space of the EM-gravity with a cosmological constant. Given the constraints \eqref{constphiVe} together with $\tilde{R}=4\Lambda$, \eqref{VphiVEexpr} and \eqref{REframe}, one can see that the curvature $R$ in the Jordan frame can be fixed to a constant value given by
	\bal
	R=R_{0}=4\Lambda e^{-\sqrt{\frac{2\kappa}{3}}\cC},\label{R0value}
	\eal
	
	From the above result, it is straightforward to see that the constant $\tilde{R}$ solution space of the EM-gravity with cosmological constant maps to the constant $R$ solution space of the $f(R)$-gravity in the Jordan frame. In the next subsection, we discuss the consistency conditions required to match the constant Ricci scalar solution space in both the EM-gravity with a cosmological constant and the $f(R)$-gravity in the Jordan frame. We will also discuss several viable forms of the $f(R)$ function that satisfy the said conditions.
	
	\subsection{Constant curvature black hole solutions for $f(R)$ gravity in Jordan frame}
	
	The action describing the $f(R)$ gravity in Jordan frame in \eqref{SJframe} where the matter contribution comes from a Maxwell field ($A_{a}$) can be given as
	\bal
	\cS_{\cJ}=\frac{1}{2\kappa}\int d^4x\sqrt{-g}\left(f(R) -\frac{1}{4}\cF^2\right),\label{SJframeF}
	\eal
	where $\cF^2=\cF_{ab}\cF^{ab}$ and $\cF_{ab}=\nabla_{a} A_{b}-\nabla_{b}A_{a}$ is the electromagnetic field strength tensor. The equations of motion for the above action are given by \eqref{SJeom} and the following Maxwell equation
	\bal
	\nabla_{a}\(\cF^{ab}\)=0,\label{JframeAeom}
	\eal
	where the traceless Maxwell stress-energy tensor is given as
	\bal
	T_{ab}=\frac{2}{\kappa}\(\cF_{ac}\cF^{cb}-\frac{1}{4}g_{ab}\cF^2\)\label{eqn:MaxwellT}
	\eal
	
	Considering the constant curvature scalar $R=R_{0}$, the trace of \eqref{SJeom} leads to
	\bal
	R_{0}=\frac{2f(R_0)}{F(R_0)}.\label{R0const}
	\eal
	which determines the curvature scalar in terms of the function $f(R)$ as long as $F(R_0)\neq 0$. The condition that the curvature scalar must assume constant real values restricts the possible form of the $f(R)$ function. This also implies the possibility that some theories of $f(R)$-gravity can give multiple real values of $R_0$ while for others one may not have a real constant value for the curvature scalar. Several models of $f(R)$-gravity, where one can have a real constant value for the curvature scalar, have been discussed in \cite{Hendi:2011eg}. Thus restricting to such theories of $f(R)$-gravity with a real constant value for the curvature scalar, one can use \eqref{R0const} in \eqref{SJeom} to obtain 
	\bal
	R_{ab}-\frac{f(R_0)}{2 F(R_0)}g_{ab}=\frac{\kappa}{F(R_0)}T_{ab}.\label{R0eom}
	\eal
	
	The above equations of motion for the $f(R)$-gravity with constant curvature scalar are reminiscent of the ones obtained for the usual Einstein gravity with a cosmological constant 
	\bal
	\Lambda=\frac{f(R_0)}{2 F(R_0)},\label{EffL}
	\eal
	and an effective gravitational constant 
	\bal
	G_{eff}=\frac{G}{F(R_0)},\label{GEff}
	\eal
	which indicates a duality between the two different theories captured by the same action in \eqref{SEframeL}. To ensure the positivity of the effective gravitational constant one has to impose the following conditions
	\bal
	F(R_0)>0,\quad F'(R_0)>0,\label{stabcond} 
	\eal
	where the second conditions $F'(R_0)>0$ is required for a stable higher-order $f(R)$-gravity \footnote{Note that the condition $F'(R_0)>0$ comes from the requirement of the positivity of, {$\left.\frac{dG_{eff}}{dR}\right|_{R=R_0}$} which ensures the stability of the $f(R)$-gravity.}. We now discuss some known models of $f(R)$ gravity with constant scalar curvature \cite{Hendi:2011eg} and discuss which of them can be described by an Einstein gravity with a cosmological constant. We also study the stability conditions for their constant scalar curvature solutions described by the following four-dimensional line element
	\bal
	ds^{2} = - g(r)\,dt^2 + \frac{ dr^2}{g(r)} + r^2\,d\Omega^2_k,\label{eq:sfr}
	\eal
	where
	\bal
	d\Omega^2_k=\begin{cases}
		d\theta^2 + \sin^{2}\theta\,d \phi^2, & k=1 \\
		d\theta^2 + d \phi^2, & k=0 \\
		d\theta^2 + \sinh^{2}\theta\,d \phi^2, & k=-1 
	\end{cases}
	\eal
	which represents the line element of a 2-sphere for $k=1$, a 2-hyperboloid $(H_2)$ for $k=-1$, and flat 2-dimensional line element for $k=0$ respectively.
	
	\subsubsection*{\textbf{Case (I): $\mathbf{f(R)=R-\mu^4/R}$ model}}
	This is one of the earliest models of $f(R)$-gravity proposed in \cite{Carroll:2003wy} to explain the positive acceleration of the expanding Universe. Interestingly, this model reduces to the usual Einstein gravity with $f(R)=R$ for very large values of the Ricci scalar. However, for small values of the Ricci scalar one can not neglect the $1/R$ term implying a modified gravity in this regime. The field equation for the Maxwell field is given by \eqref{JframeAeom} and for the metric it is given by
	\bal
	&\(1+\frac{\mu^4}{R^2}\)R_{ab}-\frac{1}{2}\(1-\frac{\mu^4}{R^2}\) R g_{ab}\nn\\
	&+\mu^4 \left(g_{ab}\Box-\nabla_a\nabla_b\right)R^{-2}=2\(\cF_{ac}\cF^{cb}-\frac{1}{4}g_{ab}\cF^2\),\label{CIfReom}
	\eal
	For the constant-curvature vacuum solutions ($\nabla_{a}R=0$), one has $R=\pm\sqrt{3}\mu^2$. Now on using the metric ansatz \eqref{eq:sfr} in \eqref{CIfReom}, one can see that the only possible solution is the Schwarzschild-AdS/dS black hole solution with
	\bal
	g(r)=k- \frac{\Lambda}{3}r^2-\frac{M}{r},\quad \Lambda=\mp\frac{\sqrt{3}}{4}\mu^2\label{CIgr}
	\eal
	which is also a solution to the Einstein gravity with the cosmological constant ($\Lambda$). Here, it should be noted that for this model of $f(R)$-gravity it is not possible to have Reisnner-Nordstrom AdS/dS black hole solutions. Furthermore, the stability conditions \eqref{stabcond} in this case become
	\bal
	&\left.\(1+\frac{\mu^4}{R^2}\)\right|_{R=\pm\sqrt{3}\mu^2}>0,\nn\\
	&\left.\(-\frac{2\mu^4}{R^3}\)\right|_{R=\pm\sqrt{3}\mu^2}>0
	\eal
	which shows that the condition $F'(\pm\sqrt{3}\mu^2)>0$ is violated for both the Schwarzschild-AdS/dS solutions. In particular, for the Schwarzschild-dS solution, this violation implies that this model suffers from the Dolgov-Kawasaki instability \cite{Dolgov:2003px}. To remove this instability from the Schwarzschild-dS solution it was proposed in \cite{Nojiri:2003ft} to add an additional $R^2$ term to the given model.
	
	\subsubsection*{\textbf{Case (II): $\mathbf{f(R)=R+\a R^n}$ model}}
	The model of $f(R)$ discussed before suffers from instability problems in the strong
	gravity/small R regime but exhibits no problems in the weak gravity/large R regime. Several viable models of $f(R)$-gravity with no instability problems in the weak gravity regime have been discussed in \cite{Nojiri:2003ft}. To resolve such instability problems in a strong gravity regime, it was proposed in \cite{kobayashi2009can} to consider the corrections proportional to higher orders of curvature such as $R^n$ for $n>1$. This is the motivation for considering the given $f(R)$-gravity model here. Solving the the field equation \eqref{SJeom} for the ansatz \eqref{eq:sfr} gives
	\bal
	&g(r)=k-\frac{\Lambda}{3}r^2-\frac{M}{r},\nonumber\\
	&k=\frac{2^n(8\Lambda)^{1-n}}{2n-4},~n\neq2,\nn\\
	&\Lambda=2^{\frac{2}{n-1}} \left(  4^n (n-2)\,\alpha\right)^{\frac{1}{1-n}}
	\eal
	which describes the metric for a Schwarzschild black hole in the presence
	of a cosmological constant. Similar to the previous case, one cannot obtain the charged solution, see \cite{kobayashi2009can, Hendi:2011eg}. The stability conditions \eqref{stabcond} in this case now reduce to
	\bal
	\frac{n-1}{n-2}>0,\quad \frac{n(n-1)}{4 \Lambda  (n-2)}>0
	\eal
	which shows that for $n<0$ the Schwarzschild-AdS solution ($\Lambda<0$) is stable. However, for $n>2$ the Schwarzschild-dS solution ($\Lambda>0$) solution is stable and free from the Dolgov-Kawasaki instability.
	
	\subsubsection*{\textbf{Case (III): $\mathbf{f(R)=R+\l \exp{\(-\xi\;R\)}}$ model}}
	An interesting and promising model of $f(R)$-gravity can be obtained by adding an exponential correction term of the form $\l \exp{\(-\xi\;R\)},\xi\in \mathcal{R}$ to the usual Einstein gravity. Notably, this model was shown to agree with cosmological observations related to the solar system and that of gravitational lensing of galaxies and clusters \cite{cognola2008class,zhang2007behavior,elizalde2011nonsingular}. Considering this form of $f(R)$-gravity in \eqref{SJeom} one gets the corresponding field equations with $T_{ab}$ as specified in \eqref{eqn:MaxwellT}. To determine the solutions, we once again substitute the ansatz \eqref{eq:sfr} for the metric in the field equations \eqref{SJeom} with the given form of the $f(R)$ function. This gives us the following
	\bal
	g(r)=k-\frac{\Lambda}{3}r^2-\frac{M}{r}+\frac{Q}{r^2},\label{expfrsol}
	\eal
	indicating that the line element in \eqref {eq:sfr} represents the Reisnner-Nordstrom AdS/dS black hole solutions. Moreover, one also has the following constraint relations
	\bal
	\Lambda=\frac{\lambda  e^{\frac{2}{Q-2}+2}}{2 (Q-2)},\quad \xi=\quad \frac{e^{-\frac{2(Q-1)}{Q-2}} (1-Q)}{\lambda},\label{expfrLxi}
	\eal
	which gives the cosmological constant $(\Lambda)$ and the parameter $\xi$ in terms of $\l$ which we consider as the only free parameter in the theory. It is straightforward to see that on setting $Q=0$ (i.e., the case of vanishing Maxwell field) one can also recover the usual Schwarzschild-AdS/dS solutions for the $f(R)$-gravity. Now to understand the stability of these solutions, we need to see in what regime the conditions \eqref{stabcond} are satisfied. In this case, the stability conditions now become
	\bal
	Q>0,\quad \frac{(Q-1)^2}{2 \Lambda  (Q-2)}>0,\label{expfrstcond}
	\eal
	which shows that Reisnner-Nordstrom AdS or dS are stable for $Q<2$ or $Q>2$ respectively. Whereas, for $Q=0$ only Schwarzschild-AdS solution is stable. 
	
	\subsubsection*{\textbf{Case (IV): $\mathbf{f(R)=R+\eta \(\log{R}\)}$ model}}
	In this case, we consider the $f(R)$-gravity model with logarithmic corrections. Such models involving the logarithm of the Ricci scalar have been studied in the past to explain the inflationary paradigm in cosmology (see \cite{Amin:2015lnh} and reference therein). Similar to the previous examples, solving the field equations \eqref{SJeom} with the choice of the given $f(R)$ for the metric ansatz \eqref{eq:sfr} one gets the following 
	\bal
	g(r)=k-\frac{\Lambda}{3}r^2-\frac{M}{r}+\frac{Q}{r^2},\label{logfrsol}
	\eal
	which represents the Reisnner-Nordstrom AdS/dS black hole solutions. These black hole solutions are accompanied by the following constraint relations
	\bal
	\Lambda=\frac{\eta}{2}\,W\left(\frac{\sqrt{e}}{2 \eta }\right),\quad Q=1+\frac{1}{2 W\left(\frac{\sqrt{e}}{2 \eta }\right)},\label{logfrLQ}
	\eal
	where $W$ stands for the Lambert-$W$ or the productlog function. The above constraints give the cosmological constant and the charge in terms of $\eta$ which is the only free parameter in the theory. For this model of $f(R)$-gravity one can also obtain the Schwarzschild-AdS/dS black hole solutions for vanishing Maxwell field. Moreover, in this case, the stability conditions \eqref{stabcond} reduce to the following 
	\bal
	\frac{2-\log{\(16\Lambda^2\)}}{1-\log{\(16\Lambda^2\)}}>0,\quad \frac{1}{\log{\(16\Lambda^2\)}-1}>0
	\eal
	which shows that all of the black hole solutions are stable as long as the condition, $2>\log{\(16\Lambda^2\)}>1$ is satisfied.
	
	In principle, one can consider a more general form of the $f(R)$ function combining the models discussed here resulting in more exotic forms of $f(R)$-gravity \cite{Hendi:2011eg,EslamPanah:2022ewo,Hendi:2012zg,EslamPanah:2022ewo}. It may also be seen that for some models of constant curvature $f(R)$-gravity only Schwarzschild-AdS/dS black hole solutions are possible whereas for others it is possible to obtain Reisnner-Nordstrom AdS/dS black hole solutions also. This observation clearly implies a duality between the constant curvature $f(R)$-gravity theories and the Einstein-Maxwell gravity with a cosmological constant. Having obtained the static spherically symmetric solutions to the cases of $f(R)$-gravity one can now look for their rotating axisymmetric stationary solutions. In general, the way to obtain such rotating solutions is to use the Newmann-Janis algorithm. However, as described in the introduction using the NJ algorithm for modified gravity theories introduces pathologies in the resulting axially-symmetric metric \cite{hansen2013applicability}. In the next section, we propose a modified NJ algorithm for obtaining rotating solutions to Einstein-Maxwell gravity with a cosmological constant. Then by exploiting the duality described here, one can show that these rotating solutions for Einstein-Maxwell gravity with a cosmological constant will also be the solutions for the constant curvature $f(R)$-gravity theories.

	\section{Newmann Janis Algorithm: Einstein gravity with cosmological constant}  \label{sec:modiNJ}
	The original Newman-Janis (NJ) algorithm was proposed as a five-step procedure for generating new rotating axisymmetric solutions from known static spherically symmetric solutions (also known as seed metric) of the Einstein equations~\cite{newman1965note,newman1965metric,Adamo:2014baa}. In this section, we describe a modified Newman-Janis algorithm that generates rotating solutions for Einstein gravity with a cosmological constant from its known static spherically symmetric solutions. Here the Schwarzschild or the  Reissner-Nordstr$\ddot{\text{o}}$m AdS/dS solutions can be considered as the seed metrices for the Einstein gravity with a cosmological constant. To describe the five steps of the said algorithm for the present case, we start with the following general spherically symmetric static seed metric 
	\begin{equation}
		ds^{2}= - F(r)\,dt^2 +  G(r)^{-1}\, dr^2 + H(r)\,\left(d\theta^2 + \sin^{2}\theta\,d \phi^2\right). \label{eq:sss}
	\end{equation}
	which is used to generate the rotating solutions. Given the above seed metric, the first step of the NJ algorithm is to write it in terms of the Eddington-Finkelstein coordinates ($x_{\mu}=\left\{u,r,\theta,\phi\right\}$) using the following transformation
	\begin{equation}
		du = dt - \frac{dr}{ \sqrt{F \,G}}. \label{eq:EF trans}
	\end{equation}
	
	The second step of the algorithm involves expressing the contravariant form of the seed metric in terms of a null tetrad, $e^{\mu}_a=\left\{l^{\mu}, n^{\mu}, m^{\mu}, \Bar{m}^{\mu}\right\}$ as:
	\begin{eqnarray}
		g^{\mu \nu} = l^{\mu} n^{\nu} + l^{\nu} n^{\mu} - m^{\mu} \Bar{m}^{\nu} -  m^{\nu} \Bar{m}^{\mu}\,, \label{eq: gmu}
	\end{eqnarray}
	where 
	\begin{eqnarray}
		l_{\mu} l^{\mu} &=& m_{\mu} m^{\mu} = n_{\mu} n^{\mu} =l_{\mu} m^{\mu} = n_{\mu} m^{\mu}= 0,\nonumber \\
		l_{\mu} n^{\mu} &=& - m_{\mu} \Bar{m}^{\mu} = 1 
	\end{eqnarray}
	with $\Bar{m}_{\mu}$ being the complex conjugate of the $m_{\mu}$ vector. For the seed metric \ref{eq:sss}, the form of the null tetrad can be obtained as 
	\begin{eqnarray}
		l^{\mu} &=& \delta_{r}^{\mu},\nonumber\\
		n^{\mu} &=& \sqrt{F / G} \,\delta_{u}^{\mu} - \left( F /2\right)\,\delta_{r}^{\mu}, \nonumber \\
		m^{\mu} &=& \left( \delta_{\theta}^{\mu} + \frac{i}{\sin{\theta}} \, \delta_{\phi}^{\mu} \right) / \sqrt{2 H}.\label{NT}    
	\end{eqnarray}
	
	Having obtained the null tetrad, the third step is to extend the Eddington-Finkelstein coordinates ($x_{\mu}$) to a new set of complex coordinates using the following transformation 
	\begin{eqnarray}
		d\tilde{u} &\to & du + i \,a\,P(\theta),\quad d\tilde{r} \to dr-i\,a\,\sin{\theta},\nonumber\\
		d\tilde{\phi} &\to& d\phi+ i\,a\,Q(\theta),\quad \theta\to \theta.\label{eq:complex}
	\end{eqnarray} 
	where $a$ is some constant and the old tetrad and metric are recovered when one imposes the constraint, $x_{\mu}=\bar{x}_{\mu}$ to the above coordinate transformation. Here it is to be noted that the usual NJ algorithm for the Einstein gravity in flat spacetime involves the complexification of only $\left\{u,r\right\}$-coordinates. However, in the present case of Einstein gravity with a cosmological constant, one requires an additional complexification of $\phi$-coordinate as well.  Thus to summarize, the effect of this transformation is to create a new metric whose components are (real) functions of the complex coordinates. For the modified NJ-algorithm being discussed her, we will follow the approach adopted in \cite{Azreg-Ainou:2014pra,Azreg-Ainou:2014aqa,Azreg-Ainou:2014nra}. On using the transformation given in eq.(\ref{eq:complex}), the components $F(r)$, $G(r) $ and $H(r)$ of the metric (\ref{eq:sss}) transform in to the new functions $A(\bar{r},a)$, $B(\bar{r},a)$ and $ C(\bar{r}, a)$ respectively. We now consider the following ansatz for the functions $A$, $B$ and $C$ 
	\begin{eqnarray}
		A(\bar{r})= A(r,\theta) &=& \frac{\Delta_{r}(r,) - a^2\,\sin^{2}{\theta}\,\Delta_{\theta}(\theta)}{r^2 + a^2\,\cos^2\theta}, \nonumber \\
		B(\bar{r})= B(r,\theta) &=& \frac{1}{A(r,\theta)},\nonumber \\
		C(\bar{r})=  C(r,\theta) &=& \frac{\left(r^2 + a^2\,\cos^2{\theta}\right) }{\Delta_{\theta}(\theta)}
	\end{eqnarray}
	which is inspired by the Kerr-AdS metric where we match its $\left\{u u\right\} $ and $\left\{\theta \theta\right\}$ components with the functions $A(r,\theta)$ and $C(r,\theta)$ respectively. 
	
	The fourth step in the algorithm is to write the transformed null tetrad using the complex coordinate transformation introduced in (\ref{eq:complex}) as
	\begin{align}
		l^{\mu} &=\delta_{r}^{\mu}, \nonumber \\
		n^{\mu} &=\delta_{u}^{\mu} - \left( A /2\right)\,\delta_{r}^{\mu},\nonumber \\
		m^{\mu} &=\frac{1}{\sqrt{2 C}}\left( \delta_{\theta}^{\mu} + i a\left(\delta_{u}^{\mu} P- \delta_{r}^{\mu}\sin{\theta} \right) + i\left(\csc{\theta}+Q\right) \delta_{\phi}^{\mu} \right) ,\label{eq:transNT}
	\end{align}
	which on using the eq. (\ref{eq: gmu}) yields the contravariant form of the transformed seed metric with the following non vanishing elements
	\begin{align}
		g^{uu} &=\frac{2\,a^2\, P^{2}(\theta)\,\Delta_{\theta}(\theta)}{\Sigma(r,\theta)},\nonumber\\
		g^{ur} &= g^{ru}= -1 - \frac{2\,a^{2}\,P(\theta)\,\sin{\theta}\,\Delta_{\theta}(\theta)}{\Sigma(r,\theta)},  \nonumber \\
		g^{u\phi} &= g^{u \phi} =  \frac{2\,a\,P(\theta)\,\left(\csc{\theta} + Q(\theta)\right)\, \Delta_{\theta}(\theta)}{\Sigma(r,\theta) }, \nonumber\\
		g^{rr} &= \frac{2\,\Delta_r(r)}{\Sigma(r,\theta)},\quad g^{\theta \theta} = \frac{2\,\Delta_{\theta}(\theta)}{\Sigma(r,\theta)}  \nonumber \\
		g^{r\phi} &= g^{\phi r} = - \frac{2\,a\,\left(1 + Q(\theta)\, \sin{\theta}\right)\,\Delta_{\theta}(\theta)}{\Sigma(r,\theta)}, \nonumber \\
		g^{\phi \phi} &= \frac{2\,\left(\csc{\theta} + Q(\theta) \right)^{2}\,\Delta_{\theta}(\theta)}{\Sigma(\theta)}\,,
	\end{align}
	where for brevity we have introduced the function $\Sigma(r,\theta )=\left(a^2 + 2\,r^2 + a^2\,\cos{2\theta}\right)$, in the above expressions. The line element corresponding to the above-transformed metric can now be given as
	\begin{multline}
		ds^2 = \frac{2 a^2 \Delta _{\theta }(\theta ) \sin ^2\theta -\Delta _r(r)}{\Sigma(r,\theta)} \, du^2  - 2\,du\, dr  \\
		+ \frac{\Sigma(r,\theta)}{2 \,\Delta _{\theta }(\theta )} \, d \theta^2 + 2\,a\, \frac{P(\theta )}{\csc{\theta} +Q(\theta)} \, dr d\phi \\
		+ \,2\,a \, \frac{2 P(\theta ) \left(\Delta _r(r)-a^2 \Delta_{\theta }(\theta ) \sin ^2{\theta}\right)-\sin{\theta}  \Sigma (r,\theta)}{\Sigma (r,\theta) (Q(\theta)+\csc{\theta} )}\, du d\phi    \\ + \frac{4 a^2 \sin \theta  P(\theta )+\frac{\Sigma(r,\theta)}{\Delta
				_{\theta }(\theta )}+\frac{4 P(\theta )^2 \left(a^4 \Delta _{\theta }(\theta ) \sin^2{\theta} -a^2 \Delta _r(r)\right)}{\Sigma(r,\theta)}}{2
			(Q(\theta)+\csc{\theta} )^2} \, d\phi^2 \label{eq:rotEF}
	\end{multline}
	which is nothing but the line element of a rotating-AdS black hole solution in the Eddington-Finkelstein coordinates.
	
	The fifth and final step of the algorithm is to go back to Boyer-Lindquist coordinates (BLC) using the following global coordinates transformations 
	\begin{equation}
		du = d t - \frac{a^2 + r^2}{\Delta_r(r)}, \, d\phi = d \phi - \frac{a\,S}{\Delta_r(r)}\,.
	\end{equation}
	where $S$ is a constant to be determined later on. The line element of the rotating-AdS metric in BLC can now be given as follows
	\begin{multline}
		ds^{2} = \frac{2 \left(a^2 \Delta _{\theta }(\theta ) \sin ^2\theta -\Delta _r(r)\right)}{\Sigma
			(r,\theta)} \, dt^{2}  \\+ \frac{\Sigma (r,\theta)}{2 \Delta _r(r)} \, dr^2 + \frac{\Sigma (r,\theta ,a)}{2 \Delta _\theta(\theta)}\,d\theta^2 \\
		- 4\,a\, \frac{\sin ^2\theta  \left(\Delta _{\theta }(\theta )\left(a^2+r^2\right) -\Delta
			_r(r)\right)}{S \,\Sigma (r,\theta )}\, dt\,d\phi\\
		+ \frac{\sin ^2\theta  \left(2 \Delta _{\theta }(\theta )\left(a^2+r^2\right)^2 -2 a^2\Delta _r(r)
			\sin ^2\theta  \right)}{S^2\,\Sigma (r,\theta)} \, d\phi^2 \label{eq: rotNJ}
	\end{multline} 
	where in deriving the above line element, we have considered that all the off-diagonal elements of the corresponding metric should vanish except its $\left\{t \phi\right\}$ component. This further gives us two constraint relations that determine the unknown functions $P(\theta)$ and $Q(\theta)$ in terms of the functions $\Delta_r$ and $\Delta_{\theta}$ as
	\begin{eqnarray}
		P(\theta) &=& \frac{\sin \theta}{\Delta_{\theta}(\theta)} \\
		Q(\theta) &=& \csc \theta \,\left( -1 + \frac{S}{\Delta_{\theta}(\theta)}\right),\label{eq:PQrels}
	\end{eqnarray}
	
	To this end, it is to be noted that the line element in (\ref{eq: rotNJ}), derived from the modified Newman Janis algorithm discussed here, has two unknown functions $\Delta_r$, $\Delta_{\theta}$ and a constant $S$. One can fix these unknowns using the equations of motion of Einstein-Maxwell gravity with a cosmological constant whose action can be given as
	\begin{eqnarray}
		{\cal S}=\frac{1}{16 \pi G}\int d^4x\sqrt{-g}\left(R-2\Lambda-\frac{1}{4}\cF^2\right),
	\end{eqnarray}
	where $G$ is the four-dimensional gravitational constant, $\Lambda$ is the cosmological constant. On solving the Einstein field equations for the rotating metric given in eq. (\ref{eq: rotNJ}), one can determine the unknown functions $\Delta_{r} (r)$, $\Delta_{\theta} (\theta)$, and the constant $S$ as 
	\begin{eqnarray}
		\Delta_{r}(r) &=& \left(a^2+r^2\right) \left(1-\frac{\Lambda r^2}{3}\right)-2 G m r +Q^2\nonumber \\
		\Delta_{\theta}(\theta) &=& 1 +\frac{\Lambda a^2}{3} \cos^{2}\theta,\quad S=1+\frac{\Lambda a^2}{3},\label{KNadsvds}
	\end{eqnarray}
	where $Q$ is the charge of the black hole and the cosmological constant is related to the AdS or dS length ($L$) as  $\Lambda=-\frac{3}{L^2}$ or  $\Lambda=\frac{3}{L^2}$ respectively. Plugging the above form of the functions $\Delta_{r} (r)$, $\Delta_{\theta} (\theta)$, and the constant $S$ in eq.(\ref{eq: rotNJ}) it is obvious to see that one gets the line element corresponding to a Kerr-Newmann-AdS/dS black hole. The mass M and angular momentum J of the Kerr-Newmann-AdS/dS black hole are related to the parameters $m$ and $a$ through the relations $M=m/\Sigma^2,~J=a m/\Sigma^2$ respectively. One can also obtain the Kerr-AdS/dS black hole solution described by \eqref{eq: rotNJ} and \eqref{KNadsvds} for $Q=0$, to the Einstein gravity with a cosmological constant by using the same algorithm.
	\onecolumngrid
	\begin{center}
		\begin{table}[h]
			\begin{tabular}{|l|l|l|l|}
				\hline
				$\mathbf{f(R)}$\textbf{gravity} & \textbf{Dual}  & \textbf{Black hole solutions (AdS/dS )} & \textbf{Dictionary with} $\mathbf{G_{eff}=G/F(4\Lambda)}$ \\
				\hline
				$\mathbf{R-\mu^4/R}$& $\mathbf{R-2\Lambda}$ & Schwarzschild and Kerr & $\Lambda=\mp\frac{\sqrt{3}}{4}\mu^2$\\
				\hline
				$\mathbf{R+\a R^n}$& $\mathbf{R-2\Lambda}$ & Schwarzschild and Kerr & $\Lambda=2^{\frac{2}{n-1}} \left(  4^n (n-2)\,\alpha\right)^{\frac{1}{1-n}}$\\
				\hline
				$\mathbf{R+\l \exp{\(-\xi\;R\)}}$ & $\mathbf{R-2\Lambda}$ & Schwarzschild and Kerr& $\Lambda=-\frac{\lambda  e}{4},\quad \xi=\frac{1}{e\lambda}$ \\
				\cline{2-4} 
				&$\mathbf{R-2\Lambda-\frac{1}{4}\cF^2}$  & Reisnner-Nordstrom and Kerr-Newmann & $\Lambda=\frac{\lambda  e^{\frac{2}{Q-2}+2}}{2 (Q-2)},\quad \xi=\quad \frac{e^{-\frac{2(Q-1)}{Q-2}} (1-Q)}{\lambda}$\\
				\hline
				$\mathbf{R+\eta \log{\(R\)}}$ & $\mathbf{R-2\Lambda}$ & Schwarzschild and Kerr& $\Lambda=\frac{\eta}{2}\,W\left(\frac{\sqrt{e}}{2 \eta }\right)$ \\
				\cline{2-4} 
				&$\mathbf{R-2\Lambda-\frac{1}{4}\cF^2}$  & Reisnner-Nordstrom and Kerr-Newmann & $\Lambda=\frac{\eta}{2}\,W\left(\frac{\sqrt{e}}{2 \eta }\right),\quad Q=1+\frac{1}{2 W\left(\frac{\sqrt{e}}{2 \eta }\right)}$\\
				\hline
			\end{tabular}
			\caption{A table showing the dictionary between the $f(R)$-gravities and their duals. The parameters $G$ and $G_{eff}$ denote the gravitational constant of $f(R)$-gravities and their duals respectively.}
			\label{FrEMGtable}
		\end{table}
	\end{center}
	\twocolumngrid
	
	Now going back to the cases of $f(R)$-gravity discussed before, for $f(R)=R-\m^4/R$ one can see that the Kerr-AdS/dS black hole solution described by \eqref{eq: rotNJ} and \eqref{KNadsvds} for $Q=0$, is also a solution to the field equations \eqref{CIfReom} for vanishing Maxwell field with the identification of the cosmological constant ($\Lambda$) with the parameter $\mu$ given in \eqref{CIgr} as before. In the Table \eqref{FrEMGtable} we summarize the dualities of different $f(R)$-gravities with constant Ricci scalar and their solutions with Einstein gravity with a cosmological constant in the presence or absence of Maxwell field.
	
	\section{Discussion and Conclusion} \label{sec:Discussion}
 In this paper, we use the conformal transformation to express the $f(R)$ gravity in the Jordan to Einstein frame. We find that constant curvature $f(R)$ gravity theories in the Jordan frame are dual to Einstein-Maxwell gravity with a cosmological constant or modification in effective gravitational constant.  We show the existence of the aforementioned duality by giving specific examples of well-known $f(R)$ gravity theories. Table \ref{FrEMGtable} shows the several $f(R)$ gravity theories, their dual and the effective gravitational constant ($G_{eff}$). We further use this fact to derive the rotating blackhole solutions for generalized $f(R)$ gravity with constant curvature. Previously, the Newman-Janis algorithm was used to generate the rotating black hole solution for Einstein and modified gravity. However, the NJ algorithm is only known to give accurate rotating spacetimes for the Einstein gravity. We then present a modified NJ algorithm to generate the axisymmetric rotating spacetimes for Einstein-Maxwell gravity with CC. Our modified NJ algorithm involves an additional complexification of $\phi$-coordinate to obtain the rotating spacetime for Einstein gravity with $\Lambda$.  This additional complexification gives the tractable form of transformed rotating metric with two unknown functions which are determined from the field equations of the gravity theory under consideration. The determination of these unknown functions in transformed metric ensures that the resulting rotating solution is indeed a solution of that particular theory.

 We have presented our results for the modified gravity theories assuming the constant curvature solutions. However, we believe that one can also map the solutions of  $f(R)$-gravity with dynamical Ricci scalar to those of the EM-gravity non-minimally coupled to a scalar field in the Einstein/Jordan frame. We also plan to explore the implications of such duality between two different gravitational theories in the context of gauge/gravity duality \cite{Maldacena:1997re}. We leave these interesting avenues for future works.
	\begin{acknowledgments}
		We acknowledge Ido Ben-Dayan for offering suggestions and encouragement. PC is supported by the postdoctoral program at the Ariel University. VK acknowledges the  postdoctoral grant of Unisa.
	\end{acknowledgments}
	\bibliography{njref}
	
\end{document}